\documentclass[journal]{IEEEtran}
\usepackage{cite}
\usepackage{balance}

\ifCLASSINFOpdf
\else
\usepackage[dvips]{graphicx}
\fi

\usepackage[cmex10]{amsmath}
\usepackage{amssymb}
\usepackage{amsmath}
\usepackage{array}
\usepackage{mdwmath}
\usepackage{graphicx}

\usepackage{mdwtab}

\usepackage{algorithm}
\usepackage{algpseudocode}

\begin{document}
\title{Intelligent Wide-band Spectrum Classifier}
\author{M. O. Mughal, Behrad Toghi, Sarfaraz Hussein and Yaser P. Fallah 
        %~\IEEEmembership{Member, IEEE} and S. Kim,~\IEEEmembership{Senior Member, IEEE}% <-this % stops a space
%\thanks{\Mark{*}Corresponding author.}%        
%\thanks{This work was supported by Institute for Information $\&$ communications Technology Promotion (IITP) grant funded by the Korean government (MSIT) (No. 2017-0-00316, Development of Fundamental Technologies for the Next Generation Public Safety Communications).}% <-this % stops a space
\thanks{Authors are with the Department of Electrical and Computer Engineering, University of Central Florida, Orlando, Fl. (e-mail: ozairmughal@ieee.org).}}% <-this % stops a space
%\thanks{Manuscript received April 19, 2005; revised August 26, 2015.}}

% The paper headers
%\markboth{Submitted to: IEEE Communication Letters, September~2018}%
%{Shell \MakeLowercase{\textit{et al.}}: Bare Demo of IEEEtran.cls for IEEE Communications Society Journals}

% make the title area
\maketitle

\begin{abstract}
We introduce a new technique for narrow-band (NB) signal classification in sparsely populated wide-band (WB) spectrum using supervised learning approach. For WB spectrum acquisition, Nyquist rate sampling is required at the receiver's analog-to-digital converter (ADC), hence we use compressed sensing (CS) theory to alleviate such high rate sampling requirement at the receiver ADC. From the estimated WB spectrum, we then extract various spectral features of each of the NB signal. These features are then used to train and classify each NB signal into its respective modulation using the random forest classifier. In the end, we evaluate the performance of the proposed algorithm under different empirical setups and verify its superior performance in comparison to a recently proposed signal classification algorithm.
\end{abstract}
%
% Note that keywords are not normally used for peerreview papers.
\IEEEpeerreviewmaketitle
\section{Introduction}
\IEEEPARstart{F}{or} a cognitive radio \cite{Mitola:PC1999}, the ability to accurately infer the radio frequency environment is critical in achieving its performance objectives. This task becomes specially challenging when the radios are scanning a wide-band (WB) frequency spectrum, which may contain multiple narrow-band (NB) emissions. The task of signal estimation for WB spectrum requires very high rate analog-to-digital converters (ADC) due to Nyquist sampling requirements. However, WB spectrum in cognitive radios is assumed to be sparsely populated by NB signals and can be estimated even with sub-Nyquist sampling thanks to the advances in compressed sensing (CS) theory \cite{Donoho}. Applications of CS for signal recovery have shown to reduce both memory and energy requirements \cite{Alam_Mughal:NGMAST2013} in addition to reduce sampling rates. Once the WB spectrum is estimated, different spectral features can be extracted from the recovered spectrum.

Lately, few studies have come out focusing on feature-based signal classification methods for WB radios \cite{Mughal:cosera2015, Mughal:globalsip2015, Kres_Mughal:JSPS2015, Mughal:bayes2018}. Authors in \cite{Mughal:cosera2015} used features which included occupied bandwidths (BW) and maximum peaks ($\text{A}_{max}$) in the power spectral densities (PSD) while in \cite{Mughal:globalsip2015}, authors used spectral correlation function for NB signal classification. In both these works, authors assume some predetermined values of these features and compared the newly retrieved feature values for NB modulation classification. In \cite{Kres_Mughal:JSPS2015}, authors demonstrated their test-bed setup which they used to scan WB spectrum using software defined radios (SDR). From this scanned WB, they extracted spectral features which included carrier frequencies ($f_c$), occupied BWs and variance of the $\text{A}_{max}$, and na\"ive Bayes classifier (NBC) to form a binary classification problem to discriminate between legitimate and potentially malicious signals. More recently, authors in \cite{Mughal:bayes2018} assumed a sparsely populated WB spectrum with multiple NB modulated signals. This WB was estimated using CS and then spectral features, such as, $f_c$, occupied BWs and $\text{A}_{max}$, were extracted for each NB signal. These features were then used to train an NBC to classify different modulated signals, which included binary phase shift keying (BPSK), binary amplitude shift keying (BASK) and quadrature phase shift keying (QPSK).   

While NBC is an optimal classifier with least implementation complexity, it takes a very strong assumption of conditional independence among features which is generally not the case in a wireless communication scenario. Furthermore, the performance of NBC is also greatly impacted if the dataset is noisy. Therefore, we employ random forest classifier (RFC) \cite{breiman2001random, kontschieder2011structured} which performs better than NBC even when the features are correlated and is robust against noise. In this paper, we significantly improve the technique introduced in \cite{Mughal:bayes2018} for NB signal classification in WB spectrum. In the first step, we estimate the WB spectrum using CS, thus reducing Nyquist sampling requirement. Then we extract key spectral features of each of NB signal from the recovered WB spectrum. These features include $f_c$, occupied BWs, $\text{A}_{max}$ and area under the PSD curve, a.k.a. energy of the signal ($\text{E}_t$). These features are then used to train the RFC and ergo for classification of different NB signals in a WB spectrum. In the end, we compare the performance of our proposed method with that of \cite{Mughal:bayes2018} and validate superior performance of our proposed algorithm under various empirical setups.

\section{System Model}
We assume a network topology such that $I$ transmitter nodes ($\text {Tx}_i$) are emitting their NB signals, where the $i$-th signal is denoted by $s_i(t)$, $i \in \{1, 2, ..., I\}$. %We have no information regarding the signals' $f_c$, BW and $\text{E}_t$. 
These transmitters are assumed to be transmitting NB signals using different modulations such as BPSK, BASK, QPSK or 32-quadrature amplitude modulation (32-QAM). Furthermore, it is assumed that the NB signals $s_i(t)$ are transmitted over different $f_c$ because they could represent emitters from different services, however, this does not ensure that they are non-overlapping in frequency-domain. The receiver station ($\text {Rx}$) is assumed to be equipped with an ideal WB antenna scanning a WB frequency spectrum and receives $I$ NB signals sparsely spread over this WB spectrum. We can express such aggregated WB signal as
\begin{equation}
r(t)=\sum_{i=1}^I  h_i(t) \star s_i(t) + w^{(\sigma)}(t) 
\label{WB_signal}
\end{equation}
where $h_i(t)$ is the channel coefficient between $\text {Rx}$ and the $i$th $\text {Tx}$, $\star$ indicates the time-domain convolution operation and $w^{(\sigma)}(t)$ illustrates additive white Gaussian noise (AWGN) having zero mean and $\sigma^2$ power spectral density. In the context of cognitive radios, it is assumed that the channel coefficients have been estimated by the receiver by using blind channel estimation techniques, such as the one detailed in \cite{Blind_ch_est}. Therefore, channel estimation is not the focus of this letter. Furthermore, it has been established by extensive spectrum occupancy measurements \cite{FCC} that a good portion of the frequency spectrum is idle due to spectrum under-utilization. This suggests that the frequency spectrum of the composite signal in (\ref{WB_signal}) is sparse. The goal of CS is to estimate the spectrum occupancy of $r(t)$ over the entire WB whose nonzero support regions concede the frequency locations and bandwidths of individual NB signals $s_i(t)$.
\section{Proposed Algorithm}
In this section, we introduce preliminaries of CS as outlined in \cite{Donoho} and \cite{Mughal:bayes2018}. After that we briefly introduce RFC and finally outline our proposed algorithm.
\subsection{Compressed Sensing}
The frequency response of the composite signal $r(t)$ can by obtained by taking an $N$-point discrete Fourier transform on (\ref{WB_signal}). Collecting the frequency-domain samples into an $N \times 1$ vector $r_f$, we can write:
\begin{equation} 
\label{fourier_domain}
\mathbf{r}_f = \sum_{i=1}^I \mathbf{\Omega}_h^{(i)} \mathbf{\Lambda}_f^{(i)} + \mathbf{w}_f
\end{equation}
where $\mathbf{\Omega}^{(i)}_h$ denotes the $N \times N$ diagonal channel matrix, i.e., $\mathbf{\Omega}^{(i)}_h=diag\left(\mathbf{h}^{(i)}_f\right)$, and $\mathbf{h}_f^{(i)}$, $\mathbf{\Lambda}_f^{(i)}$ and $\mathbf{w}_f$ are the discrete frequency-domain samples of $h_i(t)$, $s_i(t)$ and $w(t)$, respectively. The signal model in (\ref{fourier_domain}) can be generalized as follows:
\begin{equation}
\mathbf{r}_f=\mathbf{H}_f \mathbf{s}_f + \mathbf{w}_f
\label{fourier_domain_general_eq}
\end{equation}
where $\mathbf{H}_f = \left[ \mathbf{\Omega}_h^{(1)},...,\mathbf{\Omega}_h^{(I)} \right]$ denotes the channel matrix for the receiver and $\mathbf{s}_f = \left[ \left( \mathbf{\Lambda}_f^{(1)} \right)^T,...,\left( \mathbf{\Lambda}_f^{(I)} \right)^T \right]^T$ denote the aggregated signal spectrum of the transmitted NB signals. Inspecting equation (\ref{fourier_domain_general_eq}) we note that the spectrum recovery task requires to estimate $\mathbf{s}_f$ provided we have $\mathbf{H}_f$ and $r(t)$. Noting that the composite signal $r(t)$ is sparse in frequency-domain, thus we can benefit from advances in CS theory and sample our signal at sub-Nyquist rate to reduce receiver ADC complexity. 
\begin{figure}[t]
\centering
\includegraphics[width=8cm]{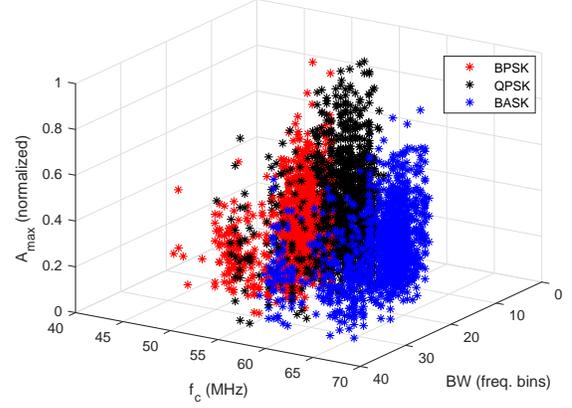}
\caption{Linearly inseparable classes for different modulated signals.}
\label{feature_space}
\end{figure}

In order to collect compressed time-domain samples at the receiver, a compressed sensing matrix $\mathbf{\Theta_{cs}}$ is required to be constructed. Thus, $M$ compressed samples are collected at the receiver in a $M \times 1$ sample vector $\mathbf{z}_t$ as shown below:
\begin{equation}
\mathbf{z}_t=\mathbf{\Theta}_{cs} \mathbf{r}_t
\end{equation}
where $\mathbf{\Theta}_{cs}$ denotes the $M \times N$, s.t. $M \leq N$ projection matrix, and $\mathbf{r}_t$ is the $N \times 1$ vector of discrete-time representations of $r(t)$. A practical design for compressive sampler that have been introduced in the literature include \cite{random_sampler:ICASSP08}.

Given that we have $M$ compressed measurements, the frequency response $\mathbf{s}_f$ in equation (\ref{fourier_domain_general_eq}) can be expressed as follows:
\begin{equation}
\mathbf{z}_t=\mathbf{\Theta}_{cs}^T \mathbf{F}_N^{-1} \mathbf{H}_f \mathbf{s}_f
+ \tilde{\mathbf{w}}_f
\label{eq_zt}
\end{equation}
where $\mathbf{r}_t = \mathbf{F}_N^{-1} \mathbf{r}_f$ and $\tilde{\mathbf{w}}_f=\mathbf{\Theta}_{cs}^T \mathbf{}{F}_N^{-1} \mathbf{w}_f$ is the white Gaussian noise sample vector. Because the emitted signals' occupancy is assumed to be low in frequency-domain in the context of cognitive radios, the composite signal spectrum $\mathbf{s}_f$ is sparse. The sparsity is indicated by $\ell$-norm measure of the signal, i.e, $||\mathbf{s}_f||_{\ell}, \ell \in [0,2)$, where $\ell=0$ indicates exact sparsity.

Because we have a sparse $\mathbf{s}_f$, equation (\ref{eq_zt}) represents a linear regression problem. By solving an $\ell_1$-regularized least squares problem at the receiver, $\mathbf{s}_f$ can be estimated as follows:
\begin{equation}
\underset{\mathbf{s}_f}{\text{min}} ||\mathbf{s}_f ||_1 +
 \lambda \left|\left|\mathbf{z}_t - \mathbf{\Theta}_{cs}^T \mathbf{F}_N^{-1} \mathbf{H}_f \mathbf{s}_f \right|\right|_2^2
\end{equation}
where the $\ell_1$-norm minimization term enforces the sparsity and $\lambda$ is a positive scalar weighting coefficient balancing the variance-bias trade-off. Equivalently, convex optimization problem can be formulated to recover the $\mathbf{s}_f$ as follows:
\begin{equation}
\mathbf{\hat{s}}_f = \text{arg} \, \underset{\mathbf{s}_f}{\text{min}}
||\mathbf{s}_f ||_1,\quad s.t. \quad
\mathbf{z}_t=\mathbf{\Theta}_{cs}^T \mathbf{F}_N^{-1} \mathbf{H}_f \mathbf{s}_f
\label{eq_estimated_sf}
\end{equation}
Various approaches are proposed in the literature for solving optimization problems such as the above mentioned equation. Among which, one can refer to the Linear Programming as in Basis Pursuit (BP) \cite{BP} or Greedy Algorithms such as Orthogonal Matching Pursuit (OMP) \cite{OMP}. The choice of recovery algorithm is of no significance for this work, therefore, we use the conventional BP algorithm for the sake of consistency with \cite{Mughal:bayes2018}. From the recovered $\mathbf{\hat{s}}_f$, spectral features of each NB signal can be extracted. Fig. \ref{feature_space} shows an example of a feature space with linearly inseparable classes for different modulation schemes.
\subsection{Random Forest Classification}
As explained earlier, the signal's features may not be independent in a wireless communications scenario, therefore, we choose RFC~\cite{breiman2001random,kontschieder2011structured} for classification as it is robust to noise and over-fitting, and works well when the features are correlated.
%can be implemented in parallel~\cite{sharp2008implementing} and have significantly faster training and testing times. 
Consider the input data represented as $\mathbf{x}=[x_1, x_2 \dots, x_n]^T\in\mathbb{R}^{n \times d}$ which consists of features from $n$ data points each having a dimension $d$. A decision tree classifier routes the input feature $x_i\in \mathbf{x}$ from the root of the tree to its leaf. The final class prediction pertaining to the feature $x_i$ can be obtained at the leaf $L(T_j(x_i))$, where $T_j$ corresponds to a tree with an index $j$. Because RFC belongs to the group of algorithms that lie within \textit{supervised learning} algorithms, therefore, it works in 2 phases. i.e., training (offline) and testing (online). Both these phases are explicitly explained in the following subsections.
\begin{figure}[t]
\centering
\includegraphics[width=\columnwidth]{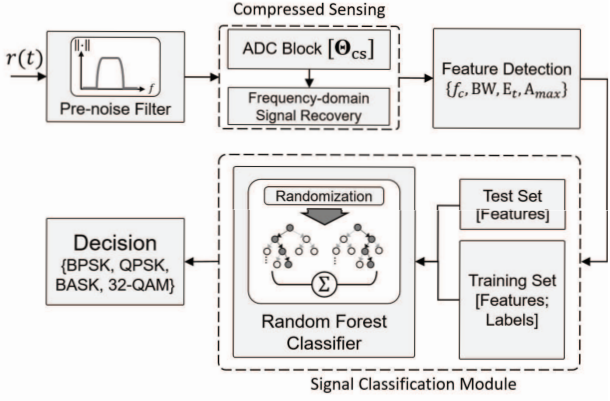}
\caption{Block diagram showing the receiver processing chain.}
\label{algo_blk}
\end{figure}
\subsubsection{Training}
Let $\mathbf{y}$ represent the set of labels such that $L(T) \in \mathbf{y}$. The other nodes $D$ of the tree are characterized by a binary decision function $\phi(x)$, which can route the feature towards right or left of the decision tree. For instance, if $\phi(x)=1$, then left sub-tree $t_l$ is selected, whereas if $\phi(x)=0$, then the data is routed towards the right sub-tree $t_r$. For a tree $T$, the prediction function can be recursively written as:
\begin{equation}
f(x|D(t_r,t_l,\phi))=\left\{\begin{matrix}
f(x|t_r) & & if\, \,  \phi(x)=0\\ 
f(x|t_l) & & if\, \,  \phi(x)=1
\end{matrix}\right.
\end{equation}
The binary decision at the nodes can be selected either randomly or by a well-defined criterion. This criterion can be modeled by a split which can optimally separate the training data. This optimality can be measured by the information gain which is represented as:
\begin{equation}
    \Delta E=-\sum _i \frac{|\textbf{x}'|}{|\textbf{x}|}E(\textbf{x}'),
\end{equation}
where $\mathbf{x}'$ is the partition from the data $\mathbf{x}$ and $|\, \cdot \, |$ is the size of the set. Moreover, $E(x')$ is the entropy which is represented as: $E(x')=-\sum_{j=k}^C p_k\, \, \log_2\, (p_k)$. Here, $p_k$ is the proportion of examples belonging to class $k \in C$, with $C$ denoting the number of classes. 
\subsubsection{Testing}
While testing, the final class label $y^*$ corresponding to a testing example $x*$ is given by:
\begin{equation}
y^*=\text{arg} \, \underset{k \in C}{\text{max}} \, \, \sum_{j=1}^U \, G(\, L(T_j)=k \,).
\end{equation}
In the above equation $U$ is the number of trees, whereas, $G(\cdot)$ is the indicator function, which is equal to 1 when $L(T(j))=k$ and 0 otherwise.  
\subsection{Proposed Algorithm}
In what follows we briefly outline the working of the proposed cognitive signal classification algorithm. 

The received signal $r(t)$ is first passed through a noise pre-filter. After that it is sampled with random sampler using $\mathbf{\Theta}_{cs}$ and estimated using BP in the frequency-domain, i.e., $\mathbf{\hat{s}}_f$. From this estimated frequency-domain WB signal, spectral features, such as, $f_c$, occupied BW, $\text{A}_{max}$ and $\text{E}_t$, are extracted for each of the NB signals. These features are then labeled with their respective modulation class, thus, building the dataset to be used for training the RFC. The trained RFC can then be fed with the unlabelled data containing features of the NB modulated signals which are needed to be classified. We show the processing chain as a block diagram in Fig. \ref{algo_blk}  for clarity of exposition.

The general time complexity for training a decision tree with $d$ samples can be stated as $\mathcal{O}(nd\log(n))$. For a forest with $U$ trees, the time complexity becomes $\mathcal{O}(Und\log(n))$. For testing, the time complexity is represented as $\mathcal{O}(U\log(n))$. 
% The space complexity of the RFC is given by $\mathcal{O}(Und)$.

% As RF classifier is based on random sampling of subspace i.e. random feature sampling as well as bagging, i.e. random sampling of examples for training, having $n>>U$, may lead to some examples being missed by the classifier. Similarly,  with $d>>U$, there exists a possibility of some features being totally omitted by all subspaces. On the other hand, $U$ being too large may put high computational strain on the classifier, which can lead to low accuracy gains with substantially higher computational costs.
%
\section{Empirical Results and Discussion}
\begin{figure}[t]
\centering
\includegraphics[width=8cm]{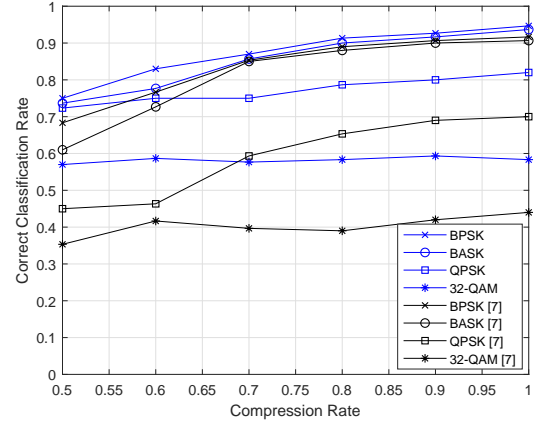}
\caption{Correct classification rate versus the compression rate at SNR = 5dB.}
\label{compression_result}
\end{figure}
We consider a WB spectrum in the range of [0,100] MHz to be under observation to conduct simulations. It was assumed that this WB was sparsely inhabited by various NB signals, such as, BPSK, QPSK, BASK and 32-QAM. Among different available recovery algorithms, BP was applied for signal recovery through CS. Compression ratios ($M/N$) were altered between 0.5 and 1.0 while the SNR was ranged from -6dB to 9dB. A total of 1500 Monte-Carlo iterations were performed to carry out the simulations. Out of which, 1200 runs were used for training the classifier and 300 runs were used for prediction. For the same empirical environment, the results of \cite{Mughal:bayes2018} are also shown with na\"ive Bayes classifier trained with the four features considered in this work for fair comparison. The classification results are generally shown with the help of confusion matrices because they clearly portray the confusion of one class with the others. However, a graphical representation is used in this work to cater for space limitations.

The rate of correct classification versus the varying compression rate at a fixed SNR of 5dB is shown in Fig. \ref{compression_result}. We can clearly observe that at Nyquist sampling rate, i.e., 200 Msamples/sec., the classification rate for BASK and BPSK is approximately 94$\%$ with the proposed algorithm while it is approximately 91$\%$ with the algorithm proposed in \cite{Mughal:bayes2018}. This performance difference becomes more noticeable as we decrease the $M/N$ to 0.5, i.e., 100 Msamples/sec., where BASK and BPSK has an approximately 75$\%$ correct classification with proposed algorithm compared to 65$\%$ correct classification with algorithm in \cite{Mughal:bayes2018}. A similar performance difference between the proposed algorithm and that of \cite{Mughal:bayes2018} is visible for QPSK and 32-QAM modulated signals where the proposed algorithm is performing better than that of ref. \cite{Mughal:bayes2018}, even at low compression rate.
%where QPSK and 32-QAM has 72$\%$ and 57$\%$ classification accuracy respectively with the proposed algorithm compared to 45$\%$ and 35$\%$ classification accuracy respectively with algorithm of \cite{Mughal:bayes2018} at $M/N=\frac{1}{2}$. 

\begin{figure}[t]
\centering
\includegraphics[width=8cm]{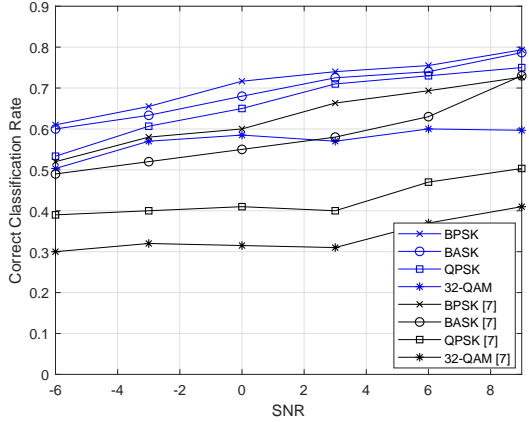}
\caption{Correct classification rate versus SNR at compression rate = $\frac{1}{2}$.}
\label{snr_result}
\end{figure}

To observe the performance of the algorithm at different SNRs, we plot the correct classification rate versus the varying SNR values at a fixed compression rate of $\frac{1}{2}$ in Fig. \ref{snr_result}. At moderately high SNR of 9dB, we can observe the classification rate of approximately 80$\%$ for BASK and BPSK with proposed algorithm while it is 73$\%$ with the algorithm in \cite{Mughal:bayes2018}. At low SNR of -6dB, the classification rate is 60$\%$ with the proposed algorithm compared to that of 50$\%$ with the algorithm in \cite{Mughal:bayes2018}, for BASK and BPSK modulated signals. Similarly, we can observe better classification performance for both QPSK and 32-QAM using proposed algorithm when compared to the algorithm proposed in \cite{Mughal:bayes2018}, over a range of SNR values even at low compression rate, further attesting the superiority of the algorithm proposed in this work.

 Although the performance difference is small for binary and quadrature modulations, it is significantly large for higher constellations, e.g., 32-QAM modulation. However, we observe sweepingly better classification performance both at low compression rates and at low SNR values for the proposed algorithm when compared to the algorithm of \cite{Mughal:bayes2018}. These observations validate the use of CS for WB spectrum recovery as the classification performance does not degrade significantly when the sampling rate is reduced. Finally note that 32-QAM classification performance is the poorest among all modulations considered, indicating that additional features are required for better classification of higher constellation modulations. These features may include cyclic frequency features or higher order statistical features. This poses an interesting research topic to study more discriminating features and advanced machine learning algorithms for higher constellation modulations classification.\balance
\section{Conclusion}
Signal classification is a classical problem in communication systems which is often carried out by demodulating the signals at the receiver. However, with the use of learning algorithms, signals can be classified without the need for demodulation. In this paper, we proposed a novel algorithm for signal classification using RFC for WB radios. To soften the very high rate sampling requirement, we used CS for sparsely populated WB spectrum. After that, we used an RFC to classify different modulated signals into their respective classes. In the end, we compared our proposed algorithm with a recently proposed signal classification algorithm to highlight the improved classification performance achieved by our proposed method. We expect that this prelusive work will inaugurate new research directions in designing cognitive receivers and securing physical layer communications.
%
%\bibliography{C:/Users/WSL_TOM/Dropbox/WORK/Bib/bibozair}{}
\bibliography{bibozair}{}
\bibliographystyle{unsrt}

\end{document}